\documentclass[journal,final,twoside]{IEEEtran}
\usepackage[pdftex]{graphicx}
\usepackage[cmex10]{amsmath}
\usepackage{epstopdf}
\usepackage{xcolor}
\usepackage{cite} 
\usepackage{flushend}
\usepackage{amssymb}
\usepackage{amsfonts}
\usepackage{multirow}
\usepackage{algorithm}
\usepackage{algorithmic}
\usepackage[flushleft]{threeparttable}
\usepackage[colorlinks=true,
            linkcolor=blue,
            urlcolor=blue,
            citecolor=blue,
            bookmarks=false]{hyperref}

\usepackage{subfigure}
\usepackage{amsmath}
\usepackage{enumitem}

\newtheorem{rem}{Remark}

\allowdisplaybreaks

\usepackage{circuitikz}

\newcommand\articletitle{Model Predictive Coolant Allocation for Integrated Tab–Surface Cooling of Battery Cells}

\begin{document}
\title{\articletitle}
\author{Godwin~K~Peprah,~\IEEEmembership{Student Member,~IEEE}, Torsten~Wik,~\IEEEmembership{Member,~IEEE}, Masood Tamadondar, and Changfu~Zou,~\IEEEmembership{Senior Member,~IEEE}    
\thanks{This work was funded in part by the Swedish Research Council (Grant No.~2023-04314) and the Swedish Energy Agency through the Swedish Electromobility Centre (Grant No.~13011).}
\thanks{Godwin K. Peprah, Torsten Wik, and Changfu Zou are with the Department of Electrical Engineering, Chalmers University of Technology, 41296 Gothenburg, Sweden (e-mail: godwinp@chalmers.se; tw@chalmers.se;  changfu.zou@chalmers.se).} 

\thanks{Masood Tamadondar is with the Energy Storage System and Electromobility Department, Volvo Group Trucks Technology, BF32420 CampX Gothenburg, Sweden (e-mail: masood.tamadondar@volvo.com).}
}

\maketitle

\begin{abstract}
Battery electrical tab cooling is effective at reducing internal thermal gradients by exploiting the high thermal conductivity of the current collectors, whereas surface cooling is effective at reducing temperature rise because of its large heat transfer area. Using either strategy alone, however, limits the achievable trade-off between thermal uniformity and temperature rise reduction. This work proposes an integrated tab-surface cooling (ITSC) system in which coolant is dynamically allocated among the lateral surface and tab channels. The allocation is formulated as an optimal control problem in which the battery temperature is regulated towards a desired reference and thermal gradients are minimised. To support this formulation, a first-principles coolant model is developed and coupled with battery and valve-actuation models. The resulting optimal coolant-allocation problem is solved using a computationally efficient real-time iteration model predictive control (RTI-MPC) scheme, with a nonlinear MPC serving as a closed-loop performance benchmark. Evaluation results under realistic driving conditions showed that RTI-MPC reproduces the nonlinear MPC thermal response with absolute errors below 0.0035\,$^\circ$C while reducing the computational cost from several seconds to 19.3~ms, indicating strong potential for real-time implementation. Additionally, evaluation of the proposed ITSC system against conventional cooling configurations demonstrates that ITSC achieves the best overall trade-off between temperature regulation and thermal gradient reduction. 
\end{abstract} 

\bstctlcite{BSTcontrol}

\section{Introduction} \label{sec: Intro}
\IEEEPARstart{T}{hanks} to their long extended cycle life, high energy and power density compared to other rechargeable batteries, Lithium (Li)-ion batteries are the dominant energy storage technology for numerous applications, including electric vehicles (EVs), stationary energy storage systems, and portable electronics \cite{choi2018advanced}. Despite these advantages, their safety, power and energy capability, charge acceptance, and lifespan are strongly dependent on temperature \cite{ma2018temperature}. Operating at elevated temperatures accelerates irreversible degradation and increases the risk of thermal runaway, while large temperature differences within a cell, known as thermal gradients, could lead to uneven ageing and localised mechanical stress within the electrodes \cite{ma2018temperature,bibin2020review}. Consequently, a sophisticated battery thermal management system (BTMS) \cite{lin2021review} is not only beneficial but essential for ensuring safe, long service life, and reliable operation of these batteries.

BTMS can be broadly categorised as passive or active. Passive systems rely on natural heat dissipation mechanisms, whereas active systems employ auxiliary devices such as fans or pumps to enhance heat transfer. Established BTMS include air cooling, liquid cooling, phase-change materials, and immersion cooling \cite{lin2021review, tete2021developments, roe2022immersion}. Air cooling is simple and inexpensive but offers limited heat removal capacity, limiting its suitability for high-power applications. Liquid cooling, typically using water-glycol mixtures, provides much higher heat transfer coefficients and is widely adopted in EVs. Phase-change materials passively buffer transient heat generation through latent heat absorption, but their low thermal conductivity and limited ability to dissipate accumulated heat can restrict long-term thermal regulation. Immersion cooling achieves a more uniform temperature distribution by directly submerging cells in dielectric fluids at the expense of increased system complexity, sealing requirements, and material compatibility constraints. Each of these approaches involves a trade-off in thermal performance, cost, weight, and complexity, driving continuous research efforts into hybrid and optimised strategies.

Within liquid-based BTMS, the design of cell-level cooling interfaces critically influences thermal behaviour. Two methods that have received particular attention are lateral surface cooling and electrical connection tab (terminals) cooling \cite{zhao2019cool}. For Li-ion pouch cells, Hunt~{\it et al.}  \cite{hunt2016surface} and Zhao~{\it et al.}  \cite{zhao2018modeling} showed that due to the larger available lateral heat transfer area, surface cooling is generally more effective at limiting the temperature rise of the cell, particularly under high current operation. 
In contrast, tab cooling removes heat directly through the current collectors, leveraging their high thermal conductivity to homogenise the temperature distribution across the electrode layers, thereby reducing internal thermal gradients. 
Dondelewski~{\it et al.} \cite{dondelewski2020role} in a follow-up study on a large-format lithium-iron-phosphate (LFP) pouch cell,
reported similar results. Under a 2C rate over 103 cycles, surface cooling resulted in a lower temperature rise and slower degradation 
whereas tab cooling yielded a higher temperature rise 
but improved thermal homogeneity and usable capacity. Similar trade-offs were reported for cylindrical cells in \cite{bolsinger2019effect, li2021optimal} where tab cooling reduced internal temperature non-uniformities 
relative to surface cooling, albeit at the cost of higher temperature rise. Complementary findings in \cite{ahmad2022identifying} further showed that surface cooling produced a faster overall temperature drop owing to the larger exposed area, while tab cooling delivered superior thermal uniformity, with peak-to-peak temperatures approximately half those observed under surface cooling in their test cell.

These studies collectively indicate that surface and tab cooling offer distinct but complementary benefits. However, each method has inherent limitations, i.e., neither strategy can simultaneously minimise temperature rise and internal gradients alone, particularly under demanding operating conditions such as fast-charging or elevated ambient temperatures. To date, 
there is 
no battery control framework in the state-of-the-art literature that systematically investigates the optimal integration of these two cooling strategies for advanced battery thermal management. 

To bridge this gap, 
we propose a control framework for integrated cooling, with the main contributions summarised as follows:
\begin{enumerate}
\item A control-oriented dynamic coolant model for multi-channel battery cooling is developed and coupled with existing battery thermal and electrical models, enabling flow-based actuation of the battery cooling process. Additionally, a valve-actuation model is developed and incorporated to capture the dynamic routing of coolant across the channels, thereby reflecting actuator-level behaviour.
\item The integration of lateral surface and tab cooling is formulated as an optimal control problem, where the available coolant flow is dynamically allocated across multiple cooling channels to regulate the cell's temperature rise toward a desired reference while suppressing internal thermal gradients. 
\item The resulting optimal coolant-allocation problem is solved for the first time within a real-time iteration model predictive control (RTI-MPC) framework to improve computational efficiency and support real-time implementation.
\end{enumerate}
The proposed control framework is systematically evaluated against conventional cooling configurations under real-world driving conditions to assess the achievable trade-offs between temperature regulation and thermal uniformity. 
Although it is demonstrated here using a cylindrical cell for clarity, the proposed framework can equally be applicable to pouch and prismatic cell formats with their appropriate thermal and coolant models. 

The remainder of this article is organised as follows. Section~\ref{Sec: battery_system_modelling} presents the modelling of the battery and coolant system. Section~\ref{Sec: MPC_formulation} details the proposed controller framework. Section~\ref{Sec: Control scheme evaluation} discusses the results. Finally, Section~\ref{Sec: Conclusions} concludes the work.

\IEEEpubidadjcol

\section{Control-oriented battery cooling system modelling} \label{Sec: battery_system_modelling}
This section presents the architecture for the proposed integrated battery cooling system and the models of the various subsystems therein. 

\begin{figure*}[tbh]
\centering
\includegraphics[width=1.7\columnwidth]{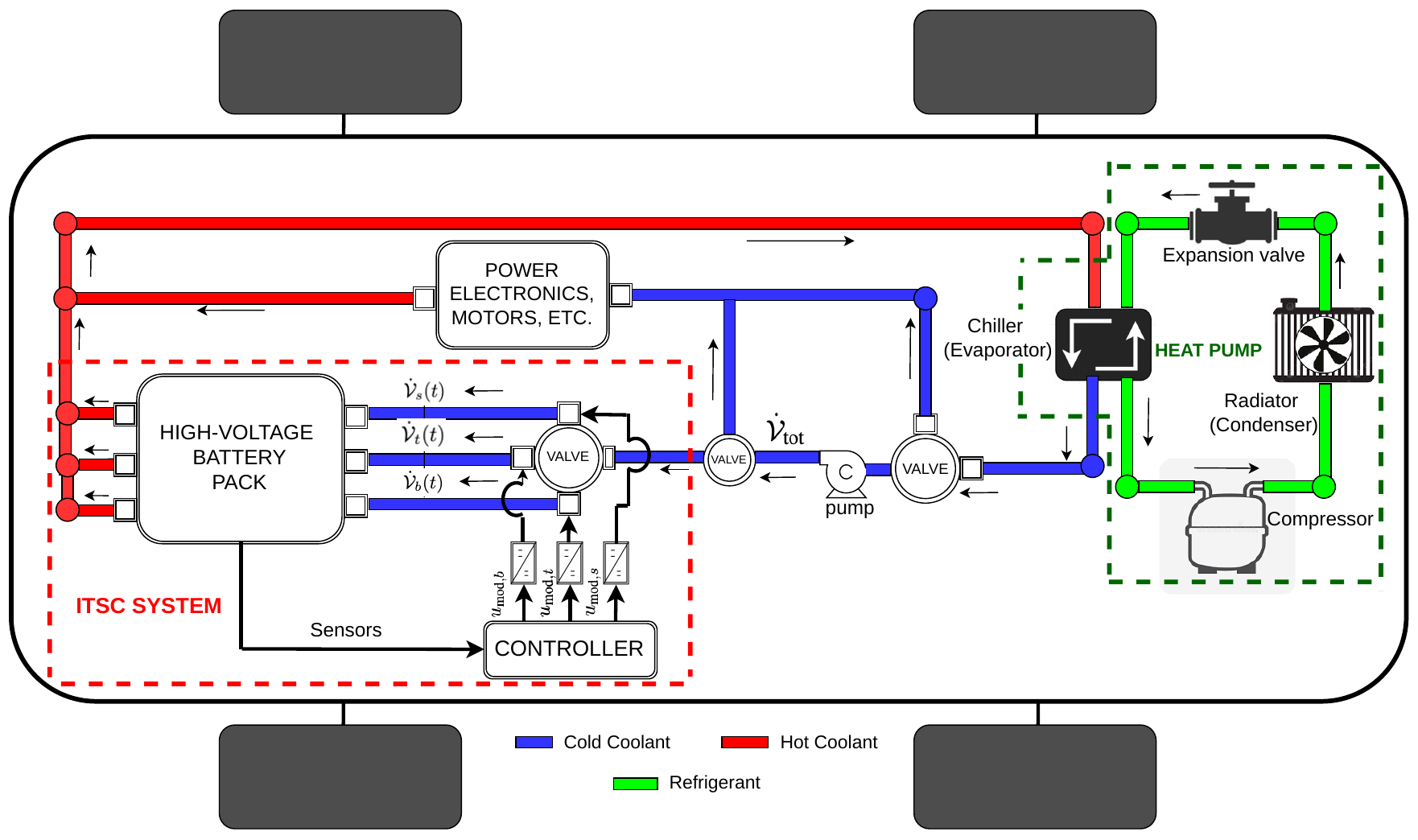}
\caption{BTMS of an EV comprising a heat pump with its refrigeration cycle (green-dashed box) and the proposed ITSC system (red-dashed box). Centrifugal pumps circulate the water-glycol mixture through the cooling loop. The red-coloured channels represent hot coolant, the blue, the cooled coolant, and the green, the refrigerant. It is assumed that the control of the chiller manages to attenuate temperature variations in the return flow from the batteries.} 
\label{Fig: EV BTMS}
\end{figure*}

Fig.~\ref{Fig: EV BTMS} shows a representative example of an EV  BTMS for a pack containing cylindrical cells. The system comprises a coolant circulation loop in which a water-glycol mixture, driven by a pump, flows through channels embedded within the high-voltage battery pack. Cooling plates attached to the curved lateral surface or base of the cells facilitate heat absorption and removal. The absorbed heat is subsequently transferred to a heat exchanger, which dissipates it to the ambient air and/or a secondary cooling circuit, such as a heat pump, depicted in the green-dashed box in Fig.~\ref{Fig: EV BTMS}. 

The proposed cooling architecture, shown in the red-dashed box of Fig.~\ref{Fig: EV BTMS} and referred to as the Integrated Tab-Surface Cooling (ITSC) system, employs three dedicated cooling channels targeting the curved lateral surface, top, and bottom sides of the cell, respectively. It has been demonstrated in \cite{li2021optimal,tranter2020communication} that cylindrical cell designs with tab areas much smaller than the top/bottom areas are not efficient for battery cooling merely from the tabs. Thus, this work considers the entire area of the top/bottom as the tab, forming the so-called all-tab or tab-less cell design. The coolant pump is rated to provide a maximum volumetric flow rate $\dot{\mathcal{V}}_{\text{tot}}$, which is distributed across the channels according to 
\begin{equation}
    \label{Eq: pump conservation}
    \dot{\mathcal{V}}_s(t) + \dot{\mathcal{V}}_t(t) + \dot{\mathcal{V}}_b(t)  \leq \dot{\mathcal{V}}_{\text{tot}}, 
\end{equation}
where $\dot{\mathcal{V}}$ denotes the volumetric flow rates with subscripts $s$, $t$, and $b$ denoting the cell's lateral surface, top, and bottom channels, respectively. This inequality 
allows partial utilisation of the pump capacity when thermal loads are low. A three-way solenoid valve \cite{taghizadeh2009modeling, angadi2022critical}, actuated via DC--DC power converters \cite{kazimierczuk2015pulse,erickson2001dc}, regulates the distribution of coolant among the channels. The volumetric flow rates $\dot{\mathcal{V}}$ are treated as the manipulated inputs, and their dynamic allocation constitutes the primary control action. For modelling purposes, coolant transport delays between the pump and the cell are neglected, as the coolant transport dynamics are significantly faster than the dominant time constant of the battery thermal dynamics. 

The proposed ITSC system integrates the battery, coolant, and valve actuation within a unified control-oriented framework, with each component modelled in the remainder of this section.


\subsection{Battery modelling}  \label{}
The battery is modelled as a coupled electrothermal system, 
as described in the following subsections.

\subsubsection{Battery thermal dynamics}  \label{}
In this subsection, we present the two-dimensional (2D) reduced-order thermal model for the cylindrical cell developed in our previous work \cite{peprah2026thermal,peprah2024control}. The governing heat conduction model and its reduced-order state-space representation via the Chebyshev spectral-Galerkin (CSG) method are briefly summarised here. They provide the cell thermal dynamics used in the subsequent coupled modelling framework.

The temperature distribution $T(r,z,t)$ within the cylindrical cell, defined over the radial coordinate $r$, axial coordinate $z$, and time $t$, is governed by the following 2D boundary value problem \cite{hahn2012heat}:
\begin{equation}
\begin{split}
    \rho_b  c_{p,b} & \frac{\partial T(r,z,t)}{\partial t} - k_{b,r} \frac{\partial^2 T(r,z,t)}{\partial r^2} \\ 
   - k_{b,r} & \frac{\partial T(r,z,t)}{r \partial r}  - k_{b,z} \frac{\partial^2 T(r,z,t)}{\partial z^2} = q(t),
    \end{split}
   \label{Eqtn: GEqn}
   \end{equation}
subject to the non-homogeneous convection boundary conditions given by
 \begin{subequations} \label{Eqtn: BC 1 2}
\begin{align}
  k_{b,r}\frac{\partial T}{\partial r} & = h_s\left(T-T_{s,\infty}\right) \:\:\:\text{ at } r = R_\text{out}, \\
   k_{b,r}\frac{\partial T}{\partial r} & = -h_c\left(T-T_{c,\infty}\right) \text{ at }  r = R_\text{in}, 
    \label{Eqtn: BC 1 2Db} \\
    k_{b,z}\frac{\partial T}{\partial z}  & = h_t\left(T-T_{t,\infty}\right) \:\:\: \: \text{ at } z = L, 
    \label{Eqtn: BC 1 2Dc}\\
   k_{b,z}\frac{\partial T}{\partial z} & = -h_b\left(T-T_{b,\infty}\right) \text{ at } z = 0,
   \label{Eqtn: BC 1 2D}
   \end{align}
\end{subequations}
 where $\rho_b$, $c_{p,b}$, $k_b$, $h$, and $T_\infty$ represent the volume-average density, volume-average specific heat capacity, thermal conductivity, convection coefficient, and fluid free-stream temperature, respectively. $R_\text{in}$ and $R_\text{out}$ are the cylindrical cell's inner and outer radius, $L$ is the length, and $q(t)$ is the heat generation rate. 
 The subscripts $s$, $c$, $t$, and $b$ denote the cell's lateral surface, core, top, and bottom, respectively. For cylindrical cells, there is no cooling in the core region, i.e., $h_c = 0$; thus, the radial temperature gradient in \eqref{Eqtn: BC 1 2Db} is zero.

Applying the CSG method to \eqref{Eqtn: GEqn}--\eqref{Eqtn: BC 1 2} yields a finite-dimensional system of coupled ordinary differential equations (ODEs) of the form  \cite{peprah2026thermal,peprah2024control}
\begin{subequations}
\label{Eq: SS-model}
\begin{align}
    \mathcal{G}_t \dot{X}_t(t) =\:& \mathcal{A}_t X_t(t) + \mathcal{B}_t U_t(t) + \mathcal{F}_t W_t(t),
    \label{Eq: state space 2D} \\
    \mathcal{Y}_t(t) =\:& \mathcal{C}_t X_t(t) + \mathcal{D}_t U_t(t), 
       \label{Eq: output equation 2D} 
\end{align}
\end{subequations}
where the state vector $X_t$ is defined as
\begin{equation}
 \label{Eqtn: state vector 2D}
X_t = \left[ c_{11},\! \cdots, c_{1N}, c_{21},\! \cdots, c_{2N},\! \cdots, c_{M1},  \!\cdots, c_{MN}\right]^{T},
\end{equation}
with the states $c_{mn}(t)$ denoting time-varying coefficients of the employed Chebyshev basis functions, $m \in \{1,\dots, M\}$, and $ n \in \{1,\dots, N\}$. The tuning parameters $M$ and $N$ denote the number of basis functions used in the $r$ and $z$ directions, respectively, and are chosen based on the desired approximation accuracy and computational budget. The resulting model order is $O=MN$. The input vector $U_t$ is composed of the cooling powers (heat flux) applied at the lateral surface, top, and bottom sides of the cell, leading to
\begin{align} 
\label{Eq: control vector 2D}
 U_t(t) = \left[ u_{s}(t), \: u_{t}(t), \: u_{b}(t) \right]^T,
\end{align}
with
\begin{subequations} 
 \label{Eq: new controls}
\begin{align} 
 u_s(t) =\:& h_s(t)\, T_{s,\infty}(t), \\ 
 u_{t}(t)=\:& h_t(t)\, T_{t,\infty}(t), \\
u_{b}(t) =\:& -h_b(t)\, T_{b,\infty}(t).
\end{align}
\end{subequations} 
The cooling powers in \eqref{Eq: control vector 2D}--\eqref{Eq: new controls} capture the thermal interaction between the battery cell and the surrounding coolant and will be explicitly related to coolant model variables in subsection~\ref{Subsec: Coolant flow dynamics modeling}. The output vector $\mathcal{Y}_t$ consists of representative temperatures within and on the cell, including the quantities required for electrothermal and battery--coolant coupling. In particular, the volume-averaged cell temperature is defined as
\begin{equation}
\label{Eq: T_v}
    T_v(t) = \frac{1}{V_b}\int_{\Omega_b} T(r,z,t)\,d\Omega_b,
\end{equation}
where $V_b$ and $\Omega_b$ are the cell volume and battery physical domain, respectively. The area-averaged battery--coolant interface temperatures at the lateral surface, top, and bottom sides are defined as
\begin{subequations}
\label{Eq: T_bc}
\begin{align}
    T_{bc,s}(t) &= \frac{1}{A_{ht,s}} \int_{A_{ht,s}} T(R_{\mathrm{out}},z,t)\,dA_{ht,s}, \\
   T_{bc,t}(t) &= \frac{1}{A_{ht,t}}\int_{A_{ht,t}} T(r,L,t)\,dA_{ht,t}, \\
   T_{bc,b}(t) &= \frac{1}{A_{ht,b}}\int_{A_{ht,b}} T(r,0,t)\,dA_{ht,b},
\end{align}
\end{subequations}
respectively, where $A_{ht}$ denotes the effective heat transfer area. The system matrices $\mathcal{G}_t$, $\mathcal{A}_t$, $\mathcal{B}_t$, $\mathcal{C}_t$, $\mathcal{D}_t$, and $\mathcal{F}_t$ are obtained via spectral-Galerkin projections of the governing heat-conduction partial differential equation (PDE) onto a finite set of basis functions, with the detailed derivations provided in \cite{peprah2026thermal,peprah2024control}. Finally, the disturbance input vector $W_t(t)$ in \eqref{Eq: state space 2D} denotes the volumetric heat generation rate $q(t)$.

\subsubsection{Battery electrical dynamics} \label{Subsec: Electrical dynamics modeling}
The battery's electrical behaviour is modelled using a lumped-parameter equivalent circuit model (ECM) \cite{hu2012comparative}, which is widely adopted for control-oriented battery modelling due to its balance between accuracy and computational efficiency. The ECM captures the dynamic current-voltage relationship of the cell. A schematic of the ECM topology is shown in Fig.~\ref{Fig: Elec_Flow_Switch}a. The model consists of an open-circuit voltage source $V_{\text{ocv}}$, which is a nonlinear function of the state of charge (SoC), a series internal resistance $R_0$, and a series connection of $p$ paralleled resistance-capacitance branches. Each branch is characterised by a resistance $R_i$ and capacitance $C_i$ to account for 
the cell's polarisation dynamics. The ECM is expressed in continuous-time state-space form as
\begin{subequations}
\label{Eq: ECM}
\begin{align}
\dot{X}_e(t) &= \mathcal{A}_e X_e(t) + \mathcal{B}_e U_e(t), \\
\hat{\mathcal{Y}}_e(t) &= \mathcal{C}_e X_e(t) + \mathcal{D}_e  U_e(t) + V_{\text{ocv}}\left(\text{SoC}(t)\right), 
\end{align}
\end{subequations}
where the state vector is defined as
\begin{equation}
X_e(t) = \left[ \text{SoC}(t), \, V_1(t), \, \cdots, \, V_p(t) \right]^T,
\end{equation}
with $V_i$ being the voltage across the $i^\text{th}$ RC branch. The input $U_e = I_b$ is the applied current, with $U_e > 0$ implying discharging and $U_e < 0$ charging. The output $\mathcal{Y}_e$ is the terminal voltage, which depends nonlinearly on SoC through $V_{\text{ocv}}$. The explicit expressions of $\mathcal{A}_e$, $\mathcal{B}_e$, $\mathcal{C}_e$ and $\mathcal{D}_e$ are provided in Appendix~\ref{App: ECM}.

\subsubsection{Battery electrothermal coupling} \label{Subsec: Electro-thermal coupling}
The heat generation rate $W_t$ that drives the thermal dynamics in \eqref{Eq: SS-model} is calculated from the thermal model output, and the electrical model states and inputs. Assuming spatially uniform heat generation, the heat generated by the cell comprises reversible component ($q_{\text{rev}}$) and irreversible component ($q_{\text{irr}}$), expressed as \cite{xie2018experimental}
\begin{subequations}
\begin{align}
q(t) = \:& \frac{q_{\text{irr}}(t) + q_{\text{rev}}(t)}{V_b}, \label{Eq: q_irrplusq_rr_Vb}\\
q_{\text{rev}}(t) =\:& U_e(t) T_v(t) \frac{dV_{\text{ocv}}}{dT_v},  \\
q_{\text{irr}}(t) =\:& U_e^2(t) R_0 + \sum_{i=1}^p \frac{V_i^2(t)}{R_i},
\end{align}
\end{subequations}
where $T_v$ is volume-averaged cell temperature defined in \eqref{Eq: T_v} and $dV_{\text{ocv}}/dT_v$ is the entropic temperature coefficient of the open-circuit voltage. 
$q_{\text{rev}}$ accounts for the entropy change associated with electrochemical reactions and $q_{\text{irr}}$ represents ohmic and charge-transfer heat generation. 
The above equations give the disturbance input in the thermal model as $W_t(t)=q(t)$.

\subsection{Coolant modelling} \label{Subsec: Coolant flow dynamics modeling}
This subsection develops a dynamic model for the bulk coolant temperature in each cooling channel, capturing the effect of coolant flow rate on convective heat transfer and advective energy transport. The model is derived from a first-principles energy balance and is formulated in a manner suitable for control design. For notational simplicity, the derivation is presented for a generic cooling channel and then applied to the lateral surface, top, and bottom channels.

Heat transfer between the battery cell and the coolant is modelled as internal forced convection, where a water-glycol mixture is circulated by a pump through cooling plates or jackets attached to the cell. This configuration is commonly used in liquid-based BTMS because it provides efficient and controllable heat removal \cite{ghasemi2024improving, zeng2023performance}. The convective heat transfer coefficient is characterised by \cite{incropera1985fundamentals}
\begin{equation}
\label{Eq: h_general}
h = \frac{\text{Nu} \cdot k_c}{D_h},
\end{equation}
where $k_c$ is the thermal conductivity of the coolant, $D_h$ is the hydraulic diameter of the flow passage, and Nu is the Nusselt number.

To derive the governing equation for the coolant temperature, we apply the first law of thermodynamics to a control volume comprising the coolant flowing through a channel in thermal contact with the battery. The following modelling simplifications are adopted:
\begin{enumerate}[label=\roman*)]
    \item \textit{Incompressible flow:} The coolant is assumed incompressible with constant density $\rho_c$ and specific heat capacity $c_{p,c}$, which can be justified by the relatively small coolant temperature variations typically encountered in BTMS operation.
    
    \item \textit{Sensible energy storage:} Only sensible thermal energy is considered, as the coolant operates below its boiling point and no phase change occurs.
    
    \item \textit{Negligible mechanical energy effects:} Kinetic and potential energy changes, as well as pressure-drop effects, are negligible relative to thermal energy transport.
    
    \item \textit{Mean temperature representation:} The coolant is characterised by a bulk mean temperature, assuming dominant streamwise heat transfer and sufficient mixing across the channel cross-section.
\end{enumerate}

Applying these simplifications to the energy balance yields the following first-order dynamical model:
\begin{equation}
\label{Eq: coolant_dynamics}
\frac{d\bar{T}_{cl}(t)}{dt} = \frac{h(t)A_{ht}(T_{bc}(t) - \bar{T}_{cl}(t))}{\rho_c V_c c_{p,c}} + \frac{\dot{\mathcal{V}}(t)(T_{cl,\text{in}}(t) - \bar{T}_{cl}(t))}{V_c},
\end{equation}
where $\bar{T}_{cl}$ is the mean coolant temperature, $T_{bc}$ is the battery--coolant interface temperature defined in \eqref{Eq: T_bc}, 
$T_{cl,\text{in}}$ is the coolant inlet temperature,  
and $V_c$ is the coolant volume in the channel. A detailed derivation of \eqref{Eq: coolant_dynamics} is provided in Appendix~\ref{App: coolant}.
The state variable is $\bar{T}_{cl}$, and the volumetric flow rate $\dot{\mathcal{V}}$ is selected as the manipulated input, as it can be directly actuated using pumps and valves in practical systems. Under this selection, system \eqref{Eq: coolant_dynamics} becomes nonlinear due to the coupling between the state $\bar{T}_{cl}$ and the control input $\dot{\mathcal{V}}$. The treatment of this nonlinearity within the control framework is detailed in Section~\ref{Sec: MPC_formulation}.

\begin{figure}[tbh]
\centering
\includegraphics[width=0.9\linewidth]{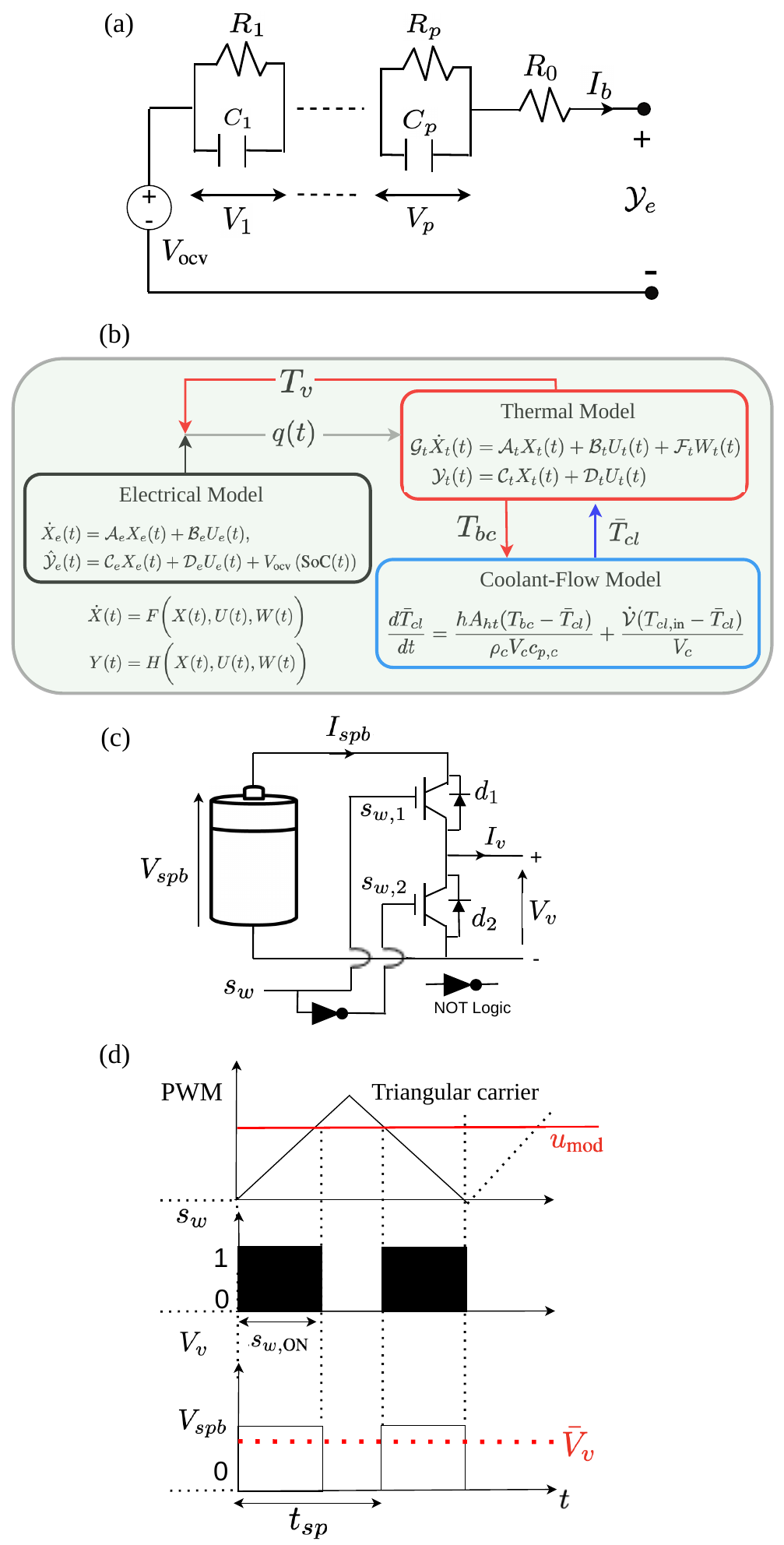}
\caption{Battery--coolant model and valve operation. (a) ECM topology (b) Interconnections among battery--coolant models. (c) A simplified schematic of a half-bridge DC-DC power converter. It consists of the switching function $s_w$, transistor switches $s_{w,1}$ and $s_{w,2}$, each with diodes $d_1$ and $d_2$ in antiparallel. (c) PWM operation showing triangular wave carrier and modulating signal. The comparison of these two signals produce the average valve voltages, $\bar{V}_{v}$ that determine the fraction of coolant flow to each channel.}
\label{Fig: Elec_Flow_Switch}
\end{figure}

\subsection{Battery--coolant coupling} \label{Sec: thermal_coolant coupling}
The battery is in thermal contact with the coolant circulating through the ITSC system. Since heat exchange occurs at the cell boundaries, the battery--coolant interaction is introduced through the thermal subsystem \eqref{Eq: SS-model}, via the boundary inputs \eqref{Eq: new controls} that govern convective heat transfer at each cooled surface. Specifically, replacing the free-stream temperatures $T_{\infty}$ in \eqref{Eq: new controls} with the corresponding mean coolant  temperatures $\bar{T}_{cl}$ yields the modified thermal boundary inputs
\begin{subequations}
\label{Eq: mod_therm_in}
\begin{align}
u_s(t) &= h_s(t)\bar{T}_{cl,s}(t), \\
u_t(t) &= h_t(t)\bar{T}_{cl,t}(t), \\
u_b(t) &= -h_b(t)\bar{T}_{cl,b}(t).
\end{align}
\end{subequations}
Equations \eqref{Eq: mod_therm_in} establish a bidirectional coupling between the battery thermal model and coolant model. The coolant model \eqref{Eq: coolant_dynamics} provides the channel mean temperatures as dynamic boundary inputs to the thermal model, while the battery boundary temperatures $T_{bc}$ computed by the thermal model drive the coolant dynamics through the convective heat transfer term in \eqref{Eq: coolant_dynamics}. The battery--coolant subsystem interconnections is illustrated in Fig.~\ref{Fig: Elec_Flow_Switch}b.

Assembling the electrical \eqref{Eq: ECM}, thermal 
\eqref{Eq: SS-model}, and coolant \eqref{Eq: coolant_dynamics} 
subsystems yields the complete nonlinear state-space model 

\begin{subequations}
\label{Eq: unified_compact}
\begin{align}
\dot{X}(t) &= F\left(X(t), U(t), W(t)\right),
\label{Eq: expanded_stateEqtn} \\
Y(t) &= H\left(X(t), U(t), W(t)\right),
\label{Eq: expanded_outputEqtn}
\end{align}
\end{subequations}
where $F(\cdot)$ and $H(\cdot)$ denote the nonlinear state and output mappings capturing the coupled battery--coolant dynamics.
The state vector is defined as
\begin{equation}
X(t) =
\begin{bmatrix}
X_e(t), & X_t(t), & X_{cl}(t)
\end{bmatrix}^T,
\end{equation}
where $X_{cl}(t)$ denotes the coolant state vector with
\begin{equation}
X_{cl}(t) = 
\begin{bmatrix}
\bar{T}_{cl,s}(t), & \bar{T}_{cl,t}(t), & \bar{T}_{cl,b}(t)
\end{bmatrix}^T.
\end{equation}
The control input vector is given by
\begin{equation}
\label{Eq: U_Vdot}
U(t) =
\begin{bmatrix}
\dot{\mathcal{V}}_s(t), & \dot{\mathcal{V}}_t(t), & \dot{\mathcal{V}}_b(t)
\end{bmatrix}^T,
\end{equation}
and the exogenous disturbance input vector is defined as
\begin{equation}
W(t) =
\begin{bmatrix}
U_e(t), & T_{cl,\text{in},s}(t), & T_{cl,\text{in},t}(t), & T_{cl,\text{in},b}(t)
\end{bmatrix}^T.
\end{equation}
The output vector is given by
\begin{equation}
Y(t) =
\begin{bmatrix}
\mathcal{Y}_e(t), & \mathcal{Y}_t(t), & X_{cl}(t)
\end{bmatrix}^T,
\end{equation}
with $X_{cl}$ assumed measurable. The mapping $F(\cdot)$ is explicitly given  by
\begin{subequations}
\label{Eq: unified_expanded}
\begin{align}
\dot{X}_e(t) &= \mathcal{A}_e X_e(t) + \mathcal{B}_e U_e(t), 
\label{eq:expanded_elec} \\
\mathcal{G}_t \dot{X}_t(t) &= \mathcal{A}_t X_t(t) + 
\mathcal{B}_t U_t\bigl(X_{cl},t\bigr) + \mathcal{F}_t W_t(t), 
\label{eq:expanded_thermal} \\
\dot{X}_{cl}(t) &= f\left(X_{cl}(t), \mathcal{Y}_t(t), U(t), T_{cl,\text{in}}(t)\right),
\label{eq:expanded_coolant}
\end{align}
\end{subequations}
where $f(\cdot)$ denotes the nonlinear coolant dynamics \eqref{Eq: coolant_dynamics}. The volumetric flow rates $\dot{\mathcal{V}}$ in \eqref{Eq: U_Vdot} constitute the effective control inputs to the coupled battery--coolant system. In practice, however, these flows are not directly actuated and are instead realised through a valve--converter actuation mechanism, described in the following subsection. 

\subsection{Valve actuation modelling}
\begin{rem}\label{Rem: valve_dynamics}
Many existing BTMS control studies treat the coolant flow rate as a directly controlled input \cite{xie2020mpc}, abstracting away the underlying actuator-level dynamics through which flow distribution is realised. As established in \cite{acker2026predictive}, this oversimplifies the coolant circuit, since the flow is an outcome of controlling physical actuators including the valves whose dynamics must be accounted for by the controller. 
\end{rem}
In response to {\it Remark~\ref{Rem: valve_dynamics}}, we explicitly model the valve actuation mechanism governing coolant distribution across the channels. For notational simplicity, the derivation is presented for a generic valve--converter pair; the resulting model applies independently to each of the three channels. 

The channel solenoid valves are assumed to be actuated using half-bridge DC--DC power converters, whose topology is illustrated in Fig.~\ref{Fig: Elec_Flow_Switch}c. The converters are powered by a supply battery providing voltage $V_{spb}$ and current $I_{spb}$. The coil voltage and current associated with each valve are denoted by $V_v$ 
and $I_v$, respectively. Each converter consists of two ideal power transistors with switching states $s_{w,1}, s_{w,2} \in \{0,1\}$. To prevent short-circuiting across the DC bus, the transistors are 
constrained by the orthogonality condition
\begin{equation}
    \int^{t}_{t-t_{sp}} s_{w,1}(\tau)s_{w,2}(\tau)
    \, d\tau = 0,
    \label{Eq: orthogonality}
\end{equation}
where $t_{sp}$ is the switching period.

The switching signal $s_w(t) \in \{0,1\}$ is generated via a pulse-width modulation (PWM) process as illustrated in Fig.~\ref{Fig: Elec_Flow_Switch}d. Specifically, a modulating signal $u_{\text{mod}}(t) \in [0,1]$, computed by the controller, is compared with a high-frequency waveform to produce the binary switching signal. 
The resulting valve voltage is given by
\begin{equation}
V_v(t) = 
\begin{cases}
V_{spb}(t), & \text{if } s_w = 1,  \\
0, & \text{if } s_w = 0.
\end{cases}
\end{equation}
To avoid explicit treatment of the high-frequency binary 
switching dynamics, we apply the switching-period averaging 
framework~\cite{kazimierczuk2015pulse, peprah2021optimal, 
peprah2025model}, which replaces the switching signals with 
their time-averaged equivalents:
\begin{equation}
\label{Eq: duty cycle}
    u_{\text{mod}}(t) := \frac{1}{t_{sp}}
    \int^{t}_{t-t_{sp}} s_w(\tau)\,d\tau 
    = \frac{s_{w,\text{ON}}}{t_{sp}}(t),
\end{equation}
where $s_{w,\text{ON}}$ is the total ON duration of $s_w$ within one PWM period. The resulting continuous-valued signals $u_{\text{mod}}$ represent the valve duty-cycles, corresponding to the fraction of time each valve remains open per switching period. The effective volumetric flow rates entering the lateral surface, top, and bottom channels are thus
\begin{equation} 
\label{Eq: controls w dc}
\dot{\mathcal{V}}_s = \dot{\mathcal{V}}_\text{tot}
u_{\text{mod},s}, \quad 
\dot{\mathcal{V}}_t = \dot{\mathcal{V}}_\text{tot}
u_{\text{mod},t}, \quad
\dot{\mathcal{V}}_b = \dot{\mathcal{V}}_\text{tot}
u_{\text{mod},b},
\end{equation}
subject to the physical duty-cycle bounds
\begin{equation}
\label{Eq: umod_bounds}
0 \leq u_{\text{mod}}(t) \leq 1,
\end{equation}
and the conservation constraint
\begin{equation} 
\label{Eq: u_conserv}
u_{\text{mod},s}(t) + u_{\text{mod},t}(t) + 
u_{\text{mod},b}(t) \leq 1,
\end{equation}
which allows the controller to utilise less than the full pump capacity when thermal loads are low. Equations~\eqref{Eq: controls w dc}--\eqref{Eq: u_conserv} complete the actuator model, closing the link between the controller outputs $u_{\text{mod}}$ and the manipulated 
inputs $\dot{\mathcal{V}}$ of model~\eqref{Eq: unified_compact}.

\section{MPC formulation} \label{Sec: MPC_formulation}
Building on the battery--coolant and valve-actuation models developed in Section~\ref{Sec: battery_system_modelling}, this section presents the control formulation used to optimally allocate coolant across the cooling channels. The control architecture,  
cost function, constraints, and controller implementation are introduced in turn.

\subsection{Control architecture and problem statement} \label{Sec: P.Statement}
The integration of lateral surface and tab cooling is formulated as the following optimal control problem: 
\emph{The optimal coolant-allocation control problem (OCAP) is to determine, over the prediction horizon, the allocation of coolant flow among the lateral surface, top tab and bottom tab cooling channels such that the battery temperature is regulated toward a desired reference  while internal thermal gradients are minimised, without violating battery safety, health-related, and actuator constraints.
} 
The regulation objective in the OCAP is motivated by the thermal degradation mechanisms that define an optimal operating window for Li-ion cells \cite{yang2018understanding, schimpe2018comprehensive, acker2026predictive, waldmann2014temperature}. Elevated temperatures accelerate side reactions such as solid electrolyte interphase (SEI) growth and electrolyte decomposition, leading to capacity fade and increased internal resistance. Conversely, lower temperatures increase the risk of Li plating during charging due to reduced ionic conductivity. These effects jointly define an optimal temperature window, motivating regulation toward a reference within this range. In addition, minimising internal thermal gradients mitigates localised ageing and thermally induced mechanical stress within the cell \cite{hunt2016surface, zhao2018modeling}.

The OCAP exhibits three key features that motivate the use of MPC. First, the battery-thermal and coolant-flow dynamics are coupled, making the coolant-allocation problem inherently multivariable. Second, the system is subject to hard constraints on battery operation and actuator limits, which must be satisfied at all times. Third, predictive optimisation enables anticipation of future thermal loads and coolant-temperature variations over the horizon. 

MPC-based BTMS have previously been demonstrated to improve temperature regulation, energy efficiency, and battery lifespan \cite{xie2020mpc, acker2026predictive}. In EV battery thermal management, RTI-MPC \cite{diehl2005real} has been adopted as a computationally efficient alternative to nonlinear (N)MPC \cite{lopez2017thermal, fischer2018demonstration}. These studies regulate a single coolant flow rate, single coolant temperature, and/or auxiliary cooling power. However, they do not address the dynamic allocation of a fixed coolant budget among complementary tab and surface cooling channels with actuator-level routing constraints. This work addresses this gap by solving, for the first time, the OCAP within an RTI-MPC framework.

\begin{figure*}[tbh]
\centering
\includegraphics[width=1.0\linewidth]{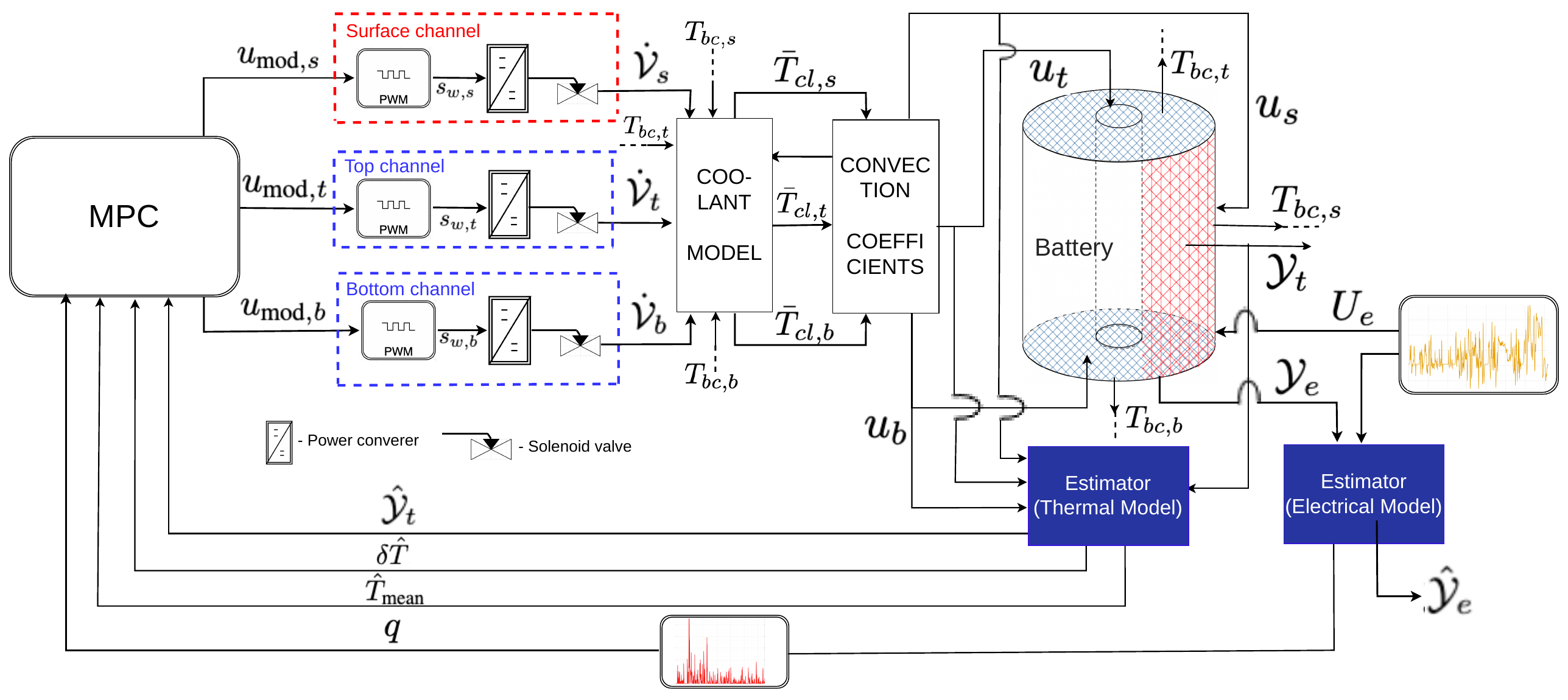}
\caption{Feedback control architecture for implementing the OCAP of a single battery cell under the proposed ITSC system. The setup comprises an MPC controller, temperature and electrical sensors, state estimation blocks, and three cooling channels for the cell's tabs and lateral surface, each governed by a valve-switching mechanism. The tabs fully cover the top and bottom areas of the cell, with their corresponding cooling channels shaded in blue. The surface cooling channel, shaded in red, covers the lateral side of the cell.} 
\label{Fig: ITSC_Feedback}
\end{figure*}

The proposed MPC-based ITSC feedback control strategy is  
illustrated in Fig.~\ref{Fig: ITSC_Feedback} for a single cell. The control architecture includes state estimation blocks for reconstructing unmeasurable battery states, particularly the SoC and internal temperatures. Based on these estimates, the controller computes optimal valve duty-cycles
to achieve the desired thermal objectives. 

\subsection{Cost function} \label{Sec: Cost function}
At each sampling instant $k$, the MPC solves the OCAP by determining the sequence of valve duty-cycles that minimise the finite-horizon objective
  \begin{equation} 
  \label{Eq: Obj}
  \begin{split}
\min_{u_{\text{mod},k}} J(k) := & \sum^{N_p-1}_{i=0} \bigg [\lambda_1 \bigg\| T^r(k+i) -  \hat{\mathcal{Y}}_t(k+i) \bigg\|^2_{Q_e} + \\ 
&\lambda_2 \bigg\| \delta \hat{T}(k+i) \bigg\|^2_{Q_\delta} + \lambda_3 \bigg\| \Delta  u_{\text{mod}}(k+i) \bigg\|^2_R \bigg],
\end{split}
\end{equation}
where $N_p$ is the prediction horizon, $\| x \|^2_{D}:=x^\top Dx$ denotes the weighted quadratic norm of vector $x$, and $\{\lambda_i \ge 0\:; i = 1,2;\; \lambda_3 > 0 \}$ are trade-off weights that determine the relative significance of each objective. The weighting matrices satisfy $Q_e \succeq 0$, $Q_\delta \succeq 0$, and $R \succ 0$. The three terms in \eqref{Eq: Obj} are described below. 

\subsubsection*{Temperature regulation}
The first term penalises the deviation of the predicted temperature output $\hat{\mathcal{Y}}_t$ from the desired reference $T^r$, which may be specified as a constant setpoint or a time-varying trajectory. This term promotes battery operation within a thermally safe and performance-optimal temperature range, necessary for mitigating temperature-induced degradation, as discussed in Section~\ref{Sec: P.Statement}. In practice, since the duty-cycle bound \eqref{Eq: umod_bounds} restricts the controller to cooling only, zero coolant flow is applied when the battery temperature $\hat{\mathcal{Y}}_t \leq T^r$.

\subsubsection*{Thermal gradient suppression}
The second term penalises internal temperature non-uniformities through the gradient-based measure $\delta \hat{T}$, defined as the root-mean-square (RMS) spatial temperature gradient of the predicted temperature field:
\begin{equation}
\label{Eq: T_gd}
\delta \hat{T}(k+i) = \left( \frac{1}{|\Omega|} \int_{\Omega} \left\| \nabla \hat{T}(r,z,k+i) \right\|_2^2 \, d\Omega\right)^{1/2},
\end{equation}
where $|\Omega|$ denotes the measure of the spatial domain $\Omega$, and $\nabla \hat{T} = \begin{bmatrix} \frac{\partial \hat{T}}{\partial r}, &  \frac{\partial \hat{T}}{\partial z} \end{bmatrix}^\top$ is the spatial temperature gradient with respect to coordinates $(r,z)$. 
Minimising~\eqref{Eq: T_gd} promotes spatially uniform temperature distributions, thereby reducing localised cell ageing and thermomechanical stress.
\subsubsection*{Control variation penalty}
The third term penalises changes in the duty-cycle input according to
\begin{equation} 
\label{Eq: slew rate}
\Delta u_{\text{mod}}(k+i) = u_{\text{mod}}(k+1+i) -  u_{\text{mod}}(k+i), 
\end{equation}
thereby discouraging aggressive changes in coolant-routing commands. This regularises the control action and reduces excessive chattering 
of the valve--converter system.

\subsection{Constraints}
The objective \eqref{Eq: Obj} is minimised subject to the following constraints $\forall i \in \{0,1,\cdots,N_p-1\}$;
\begin{subequations} 
\label{Eq: constraints}
    \begin{align}
    \hat{X}_{e}(k+1+i) &= \mathcal{A}_{e,d}\hat{X}_{e}(k+i) + 
    \mathcal{B}_{e,d} U_{e}(k+i), \label{Eq: discrete Elec_mdl}\\ 
    \mathcal{G}_t\hat{X}_{t}(k+1+i) &= \mathcal{A}_{t,d}\hat{X}_{t}(k+i) + 
    \mathcal{B}_{t,d} U_{t}(k+i) \nonumber \\
    &\quad + \mathcal{F}_{t,d}W_{t}(k+i), \\
    \hat{\mathcal{Y}}_{t}(k+i) &= \mathcal{C}_{t}\hat{X}_{t}(k+i) + 
    \mathcal{D}_{t} U_{t}(k+i), \label{Eq: discrete therm mdl}\\
    X_{cl}(k+1+i) &= f_d \bigg(X_{cl}(k+i), \hat{\mathcal{Y}_t}(k+i), U(k+i), \label{Eq: discrete flow mdl} \nonumber \\
   &\quad T_{cl,\text{in}}(k+i) \bigg), \\
    0 &\le \hat{\text{SoC}}(k+i) \leq 1, \label{Eq: elect. states limits}\\
    \hat{T}_c(k+i) &\leq \hat{T}_{c,\text{max}}, \label{Eq: core limit} \\
    U_{e,\text{min}} &\le U_e(k+i) \leq U_{e,\text{max}}, \label{Eq: elect. input limit} \\
    0 &\le u_{\text{mod}}(k+i) \le 1, \label{Eq: u_mod limit}\\
    u_{\mathrm{mod},s}(k+i) &+ u_{\mathrm{mod},t}(k+i) + 
    u_{\mathrm{mod},b}(k+i) \le 1. \label{Eq: duty sum limit}
    \end{align}
\end{subequations}
Equations \eqref{Eq: discrete Elec_mdl}--\eqref{Eq: discrete flow mdl} define the discrete-time prediction model used by the MPC, obtained by zero-order hold discretisation of the continuous-time system in \eqref{Eq: unified_expanded}. The matrices with subscript $d$ denote the discrete-time battery subsystem, while $f_d(\cdot)$ denotes the discrete-time coolant subsystem, which retains the nonlinear structure of \eqref{Eq: coolant_dynamics}. Here, ($ \: \hat{\cdot}\: $)  denotes variables 
that are either estimated by the observer or propagated forward 
through the prediction model. 

The remaining constraints enforce admissible bounds on the system states and inputs. Constraint \eqref{Eq: elect. states limits} restricts the battery SoC to the physically admissible interval $[0, 1]$, while \eqref{Eq: core limit} and \eqref{Eq: elect. input limit} impose safety limits on the estimated core temperature $\hat{T}_c$ and battery current $U_e$, respectively. Restricting $\hat{T}_c$, which often corresponds to the hottest region of the cell, mitigates the risk of accelerated degradation and thermal runaway. Finally, \eqref{Eq: u_mod limit} and \eqref{Eq: duty sum limit} enforce the physical limits of the valve--converter actuation and ensure that the total commanded coolant distribution does not exceed the available pump capacity.

\subsection{Controller Implementation} \label{Sec: Controller_Implementation}
At each sampling instant $k$, the constrained optimisation problem 
\eqref{Eq: Obj}--\eqref{Eq: constraints} is solved in a receding-horizon 
fashion. The optimal duty-cycle sequence 
$\{u^*_{\mathrm{mod}}(k+i)\}_{i=0}^{N_p-1}$ is computed, only the first 
element $u^*_{\mathrm{mod}}(k)$ is applied to the valve--converter 
actuation system, and the optimisation is repeated at the next sampling 
instant using updated state information. 

In the NMPC formulation, the prediction model \eqref{Eq: discrete Elec_mdl}--\eqref{Eq: discrete flow mdl} is retained in its original form, including the nonlinear coolant dynamics. The resulting optimisation problem is therefore a nonlinear programme, which is solved directly at each sampling instant $k$. Solving OCAP \eqref{Eq: Obj}--\eqref{Eq: constraints} to convergence at every sampling instant via NMPC can be computationally demanding, which may hinder real-time implementation. To reduce this burden, the proposed controller is implemented within the RTI-MPC scheme \cite{diehl2005real, gros2020linear}. The NMPC is used as a benchmark for assessing the closed-loop thermal performance attainable under the proposed ITSC system.

\subsubsection*{RTI-MPC for the coolant-allocation problem}
 At each sampling instant $k$, the RTI-MPC implementation consists of a warm-start step, a preparation phase, and a feedback phase. 
 \paragraph{Warm-start step} 
 Let
\begin{equation}
u^{*,\mathrm{prev}}_{\mathrm{mod}}(i), \qquad i=0,\ldots,N_p-1, 
\end{equation}
denote the optimal duty-cycle sequence computed at the previous sampling instant $k-1$. At $k$, this sequence is shifted forward by one step to initialise the nominal duty-cycle input sequence $\bar{u}_{\mathrm{mod}}$ in the current prediction horizon:
\begin{equation}
\bar{u}_{\mathrm{mod}}(k+i)
=
u^{*,\mathrm{prev}}_{\mathrm{mod}}(i+1),
\qquad i=0,\ldots,N_p-2.
\end{equation}
The terminal input is held constant as
\begin{equation}
\bar{u}_{\mathrm{mod}}(k+N_p-1)
=
u^{*,\mathrm{prev}}_{\mathrm{mod}}(N_p-1).
\end{equation}
The nominal volumetric flow trajectory $\bar{U}(k+i)$ is then obtained from \eqref{Eq: U_Vdot} through \eqref{Eq: controls w dc}. The corresponding nominal mean coolant temperature trajectory is obtained by forward propagation 
of nonlinear model \eqref{Eq: discrete flow mdl} as
\begin{equation}
\label{Eq: nominal trajectory}
\bar{X}_{cl}(k+1+i) = f_d\bigl(\bar{X}_{cl}(k+i), 
\bar{\mathcal{Y}}_t(k+i), \bar{U}(k+i), T_{cl,\text{in}}(k+i)\bigr).
\end{equation}
for $i=0,\ldots,N_p-1$, where $\bar{\mathcal{Y}}_t$ denotes the nominal battery temperature output trajectory.

\paragraph{Preparation phase}
The coolant dynamics \eqref{Eq: discrete flow mdl} are linearised about the nominal coolant trajectory \eqref{Eq: nominal trajectory}. Define the perturbation variables
\begin{subequations}
    \begin{align}
        \Delta X_{cl}(k+i) &= X_{cl}(k+i)-\bar{X}_{cl}(k+i), \\
        \Delta U(k+i) &= U(k+i)-\bar{U}(k+i), \\
         \Delta \hat{\mathcal{Y}}_t(k+i) &= \hat{\mathcal{Y}}_t(k+i)-\bar{\mathcal{Y}}_t(k+i).
    \end{align}
\end{subequations}
A first-order Taylor approximation of \eqref{Eq: discrete flow mdl} gives
\begin{align}
\label{Eq: RTI_linearised_coolant}
\Delta X_{cl}(k+1+i) &= A_{cl,i}\Delta X_{cl}(k+i) + B_{cl,i}\Delta U(k+i) \nonumber \\ &\quad 
+ E_{cl,i}\Delta \hat{\mathcal{Y}}_t(k+i),
\end{align}
where the Jacobians are evaluated along the nominal trajectory as
\begin{subequations} 
\label{}
\begin{align}
A_{cl,i} = \left. \frac{\partial f_d}{\partial X_{cl}} \right|_{\bar{\xi}_i}, \quad
B_{cl,i} = \left. \frac{\partial f_d}{\partial U} \right|_{\bar{\xi}_i}, \quad E_{cl,i} = \left. \frac{\partial f_d}{\partial \hat{\mathcal{Y}}_t} \right|_{\bar{\xi}_i}, 
\end{align}
\end{subequations} 
with $\bar{\xi}_i = \left( \bar{X}_{cl}(k+i), \bar{\mathcal{Y}}_t(k+i), \bar{U}(k+i), T_{cl,\mathrm{in}}(k+i) \right)$. 
Replacing \eqref{Eq: discrete flow mdl} with \eqref{Eq: RTI_linearised_coolant} yields a local linear approximation of the coolant model, which transforms the original OCAP into the following local quadratic programme (QP) at each $k$:
\begin{equation}
\label{Eq: RTI_QP}
\begin{aligned}
 \min_{\Delta u_{\mathrm{mod},k}} \quad
 &J(k) \\
\mathrm{s.t.} \quad & \eqref{Eq: discrete Elec_mdl}\text{--}\eqref{Eq: discrete therm mdl}, \eqref{Eq: RTI_linearised_coolant},\eqref{Eq: elect. states limits}\text{--}\eqref{Eq: elect. input limit} \\
& 0 \leq \bar{u}_{\mathrm{mod}}(k+i)
+\Delta u_{\mathrm{mod}}(k+i) \leq 1,\\
& \bar{u}_{\mathrm{mod},s}(k+i) + \Delta u_{\mathrm{mod},s}(k+i) \: + \\
& \bar{u}_{\mathrm{mod},t}(k+i) + \Delta u_{\mathrm{mod},t}(k+i) \: + \\
& \bar{u}_{\mathrm{mod},b}(k+i) + \Delta u_{\mathrm{mod},b}(k+i)
\le 1,  \\
&i = 0,\ldots,N_p-1.
\end{aligned}
\end{equation}
The QP decision variable is the input correction $\Delta u_{\mathrm{mod}}$, so that the updated duty-cycle sequence is
\begin{equation}
u_{\mathrm{mod}}(k+i) = \bar{u}_{\mathrm{mod}}(k+i) + \Delta u_{\mathrm{mod}}(k+i).
\end{equation}

\paragraph{Feedback phase} 
Once the current states $\hat{X}_e(k)$, $\hat{X}_t(k)$, and $X_{cl}(k)$ become available, QP \eqref{Eq: RTI_QP} is solved once.
The first duty-cycle command is then updated as
\begin{equation}
u^*_{\mathrm{mod}}(k) = \bar{u}_{\mathrm{mod}}(k) + \Delta u^*_{\mathrm{mod}}(k),
\label{Eq: RTI_update}
\end{equation}
and applied through \eqref{Eq: controls w dc}. The remaining elements of the updated input sequence are stored and used to warm-start the optimisation at the next sampling instant $k+1$.

This single-QP-per-step structure is the key computational advantage of RTI-MPC over the NMPC. Rather than repeatedly solving the nonlinear programme to convergence, RTI-MPC performs a single linearisation of the coolant dynamics per step and solves one structured QP, retaining the receding-horizon and constraint-handling structure of the OCAP.
The nonlinear coolant dynamics are preserved in the nominal trajectory propagation \eqref{Eq: nominal trajectory}, while the QP solves a local correction about this trajectory. The accuracy of this approximation improves as the warm-start converges toward the true optimum.

\section{Results and discussion} \label{Sec: Control scheme evaluation}

\subsection{Specification and simulation setup} \label{Sec: setup}
\begin{figure}[tbh]
\centering
\includegraphics[width=1\columnwidth]{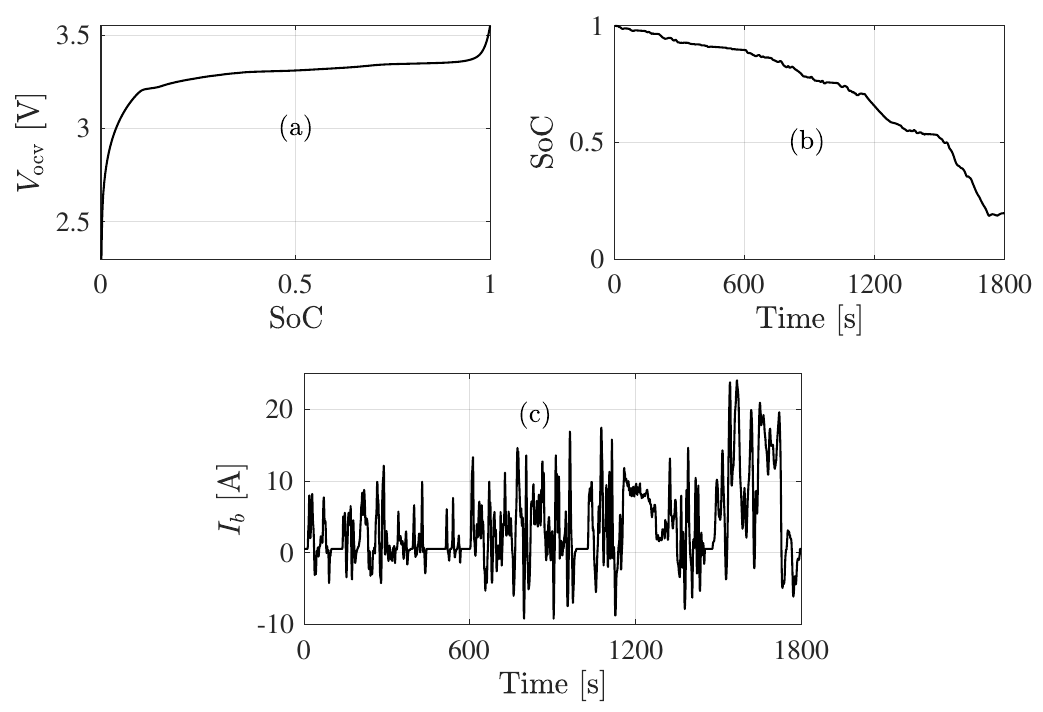}
\caption{ Electrical model variables. (a) $V_\text{ocv}$--SoC relationship for the considered LFP cell. (b) SoC trajectory under the applied mWLTP current input. (c) Applied mWLTP current input.}
\label{Fig: OCV_SOC_WLTP}
\end{figure}

The proposed ITSC system is simulated on an A123 2.3~Ah LFP cylindrical cell \cite{duh2020comparative}. The thermo-physical parameters used in the thermal model are adopted from \cite{peprah2026thermal}, namely $\rho$ = 2118~kgm$^{-3}$, $c_{p,b}$ = 795~Jkg$^{-1}$K$^{-1}$, $k_r$ = 0.67~Wm$^{-1}$K$^{-1}$, and $k_z$ = 66.6~Wm$^{-1}$K$^{-1}$. For the ECM, $R_0$ = 0.0106~$\Omega$, and a single RC branch is employed with $R_1$ = 0.0169~$\Omega$ and $C_1$ = 2249~F \cite{mohan2015estimating}. Fig~\ref{Fig: OCV_SOC_WLTP}(a) shows the nonlinear $V_\text{ocv}$--$\text{SoC}$ relationship, obtained from experimental data and implemented as a lookup table for the considered LFP cell. 

In this study, only the irreversible heat generation component $q_{\text{irr}}$ is considered. This assumption is justified by the fact that for LFP chemistry under moderate-to-high current operation, irreversible losses dominate the total heat generation, while the reversible contribution is comparatively small \cite{xie2018experimental}. 
Accordingly, the heat generation term in \eqref{Eq: q_irrplusq_rr_Vb} is approximated as $q \approx q_{\text{irr}}/V_b$.
The applied battery current $I_b$ depends on the vehicle operating conditions. We select the Worldwide Harmonised Light Vehicle Test Procedure (WLTP) \cite{mock2014wltp}, which captures a wide range of driving conditions on urban, suburban, and highway roads, to generate a realistic, dynamically varying current load. To obtain a thermally demanding scenario within the simulation window, the WLTP current trajectory is scaled by a factor chosen to maximise the mean C-rate while avoiding cell undervoltage and excessive depletion. The modified WLTP (mWLTP) has a mean C-rate of 2. 
The SoC evolution and applied current trajectory are shown in Fig.~\ref{Fig: OCV_SOC_WLTP}(b)--(c).

For the coolant, a typical 50:50 ethylene-glycol-water mixture at 20~$^\circ$C is adopted, with properties $\rho_c$ = 1069~kgm$^{-3}$, $c_{p,c}$ = 3323~Jkg$^{-1}$K$^{-1}$, and $k$ = 0.3892~Wm$^{-1}$K$^{-1}$ \cite{kim2006battery}. The convection coefficients $h$ used in equations \eqref{Eq: coolant_dynamics} and \eqref{Eq: mod_therm_in} are derived in Appendix~\ref{App: h_coeff}. Since a high heat transfer area-to-volume ratio is desirable, the channel thicknesses are chosen to be very thin: $\epsilon_s$ = 2~mm for the lateral surface channel and, for simplicity, 
$\epsilon_t = \epsilon_b$ = 2~mm for the top and bottom channels. Substituting these dimensions into \eqref{Eq: hs_t_b} and assuming laminar fully developed flow with Nu = 4.86 \cite{incropera1985fundamentals} yields the convection coefficients $h_s \approx h_t \approx h_b \approx$ 480~Wm$^{-2}$K$^{-1}$. These values are consistent with those typically reported for forced convection liquid cooling in BTMS  \cite{tete2021developments,wu2019critical}. 

The total coolant flow rate for the cell, $\mathcal{\dot{V}}_{\text{tot}}$, is sized from an energy balance based on the cell's expected thermal load. To dissipate a representative peak cell heat generation $q \approx$ 100~W with an allowable coolant temperature rise of $\Delta T_{cl} \approx$ 5~K, the required flow rate is 
\begin{equation}
   \mathcal{\dot{V}}_{\text{tot}} = \frac{q}{\rho_c  c_{p,c}\Delta T_{cl}} \approx 5.63 \times 10^{-6} \: \text{m}^{3}\text{s}^{-1},
\end{equation}
which corresponds to a nominal flow rate of approximately 0.34~Lmin$^{-1}$. 

For the CSG thermal model \eqref{Eq: SS-model}, the number of basis functions in the radial and axial directions is chosen equal, i.e., $M=N$, yielding a model order of $O=N^2$. The MPC prediction model uses $N=2$ (four states) for both the RTI-MPC and NMPC implementations, whereas the plant model uses $N=10$ (100 states).
The initial cell temperature distribution $T(r,z,0)$, initial coolant mean temperatures $X_{cl}(0)$, and inlet coolant temperatures $T_{cl,\text{in}}$ are all set to 30~$^\circ$C. 
All simulations are conducted on a MacBook Pro with a 2.6GHz 6-core Intel Core i7 processor and 16GB RAM. The optimisation problem is modelled with Yalmip \cite{lofberg2004yalmip} in MATLAB R2025b and solved using Gurobi 
for the RTI-MPC and fmincon for the NMPC. 
The MPC sampling time is set to 1~s, with a prediction horizon of $N_p$ = 5 steps. In the objective function, the temperature reference is set to $T^r$ = 35~$^\circ$C, corresponding  approximately to the midpoint of the safe and degradation-conscious battery operating temperature range highlighted in 
\cite{yang2018understanding, schimpe2018comprehensive, acker2026predictive, waldmann2014temperature}. The trade-off weights are $\lambda_1 = \lambda_2$ = 1 and $\lambda_3$ = 0.5, 
and the core temperature constraint in \eqref{Eq: core 
limit} is set to $T_{c,\text{max}}$ = 50~$^\circ$C.

\begin{figure*}[tbh]
\centering
\includegraphics[width=0.8\linewidth]{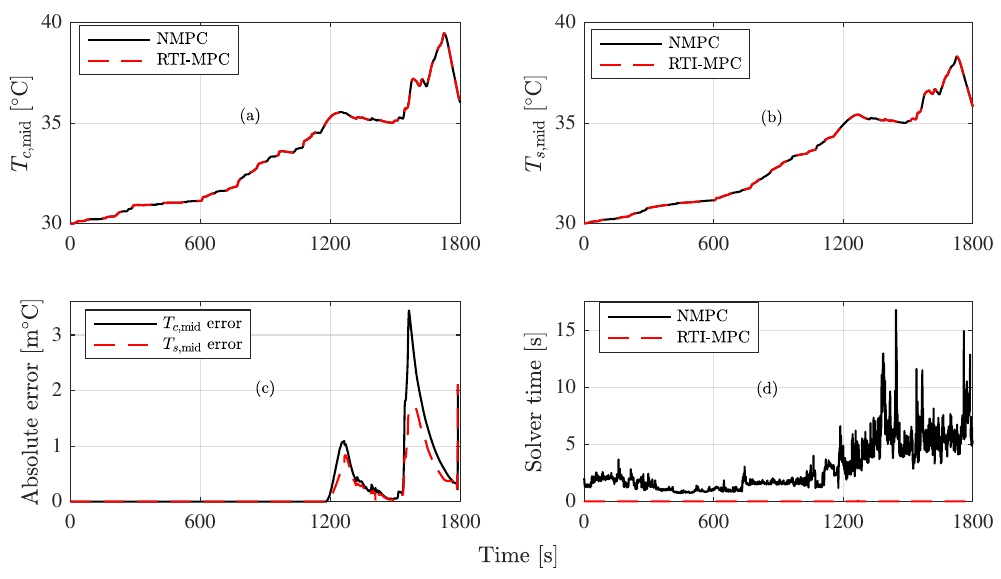}
\caption{Comparison of the NMPC and RTI-MPC implementations. (a) and (b) show the battery core mid-point and surface mid-point temperature evolutions. (c) Absolute error in core and surface mid-point temperatures between NMPC and RTI-MPC. (d) Online solver time per sampling instant.}

\label{Fig: NMPC_vs_RTI}
\end{figure*}

\subsection{RTI-MPC vs. NMPC} \label{}
The closed-loop temperature responses of the NMPC and RTI-MPC implementations are compared in Fig.~\ref{Fig: NMPC_vs_RTI}. As shown in Fig.~\ref{Fig: NMPC_vs_RTI}(a)--(b), the RTI-MPC reproduces the core and surface temperature trajectories of the NMPC almost exactly over the entire mWLTP. The corresponding absolute temperature errors in Fig.~\ref{Fig: NMPC_vs_RTI}(c) remain below 0.0035~$^\circ$C for the core and 0.002~$^\circ$C for the surface throughout the entire cycle, with only minor deviations appearing during the most thermally demanding part of the drive cycle. 
These errors are negligible relative to the temperature scales of practical interest, confirming that the successive linearisation employed in RTI-MPC introduces no meaningful loss of control performance under this operating profile.

The computational advantage of the RTI-MPC is evident from Fig.~\ref{Fig: NMPC_vs_RTI}(d). While the NMPC exhibits substantially larger online solve times and occasional spikes, the RTI-MPC maintains consistently low solver times throughout the cycle. For the present case, the RTI-MPC achieved a maximum and mean online solve times of 0.41~s and 0.0193~s, respectively, on a laptop-class processor in MATLAB, which is well below the 1~s sampling interval and indicates strong potential for real-time embedded controller implementation.

\subsection{Benchmarks for the ITSC system}\label{sec:benchmark}
To evaluate the proposed ITSC-based control strategy, its performance is compared against five baseline cooling configurations. Conventional cooling configurations for cylindrical cells apply cooling along the lateral surface only, through a single tab only, through a combination of lateral surface and single tab, or through both the top and bottom tabs \cite{worwood2018study,uwalaka2024review}. These four configurations are denoted here as surface cooling (SC), bottom tab cooling (bTC), bottom tab and surface cooling (bTSC), and bottom and top tabs cooling (btTC). A fifth  
benchmark, denoted as equal-split (ES), corresponds to a fixed uniform distribution of coolant across all available channels. This benchmark represents a predefined coolant manifold without active feedback control or optimisation, reflecting rule-based designs commonly used for their simplicity and low implementation cost.

Performance is assessed using metrics that quantify the extent to which the cell temperature overshoots 
the desired reference, temperature rise, and thermal non-uniformity. The reference overshoot is evaluated using the local tracking error
\begin{equation}
\label{Eq: error_T_Tr}
e(r,z,t)
=
\max\left(0,\,T(r,z,t)-T^r\right),
\end{equation}
with the corresponding maximum and mean deviation errors defined as
\begin{subequations}
\label{Eq: deviation_errors}
\begin{align}
e_{\max}(t) 
&= \max_{(r,z)\in\Omega} e(r,z,t), \\
e_{\mathrm{mean}}(t) 
&= \frac{1}{|\Omega|}
\int_{\Omega} e(r,z,t)\, d\Omega .
\end{align}
\end{subequations}
Since cooling action is not activated when $T(r,z,t) \leq T^r$,
temperatures below the reference do not necessarily indicate poor cooling performance; the relevant tracking error is the local temperature overshoot above $T^r$, as reflected in the one-sided definition in \eqref{Eq: error_T_Tr}. 
The temperature rise metrics are the maximum and mean cell temperatures,
\begin{subequations}
   \begin{align}
T_{\max}(t) &= \max_{(r,z) \: \in \: \Omega} T(r,z,t), \\
T_{\mathrm{mean}}(t) &= \frac{1}{|\Omega|}\int_{\Omega} T(r,z,t)\, 
d\Omega, 
   \end{align} 
\end{subequations}
and the thermal-uniformity metrics are the maximum and RMS magnitudes of the spatial temperature gradient,
\begin{subequations}
\label{Eq: gradient_metrics}
\begin{align}
\delta T_{\max}(t) 
&= \max_{(r,z)\: \in \: \Omega} \left\|\nabla T(r,z,t)\right\|_2, \\
\delta T_{\mathrm{rms}}(t) &= \left( \frac{1}{|\Omega|} \int_{\Omega} \left\|\nabla T(r,z,t)\right\|_2^2\,d\Omega \right)^{1/2}.
\end{align}
\end{subequations}

With the exception of ES, the proposed ITSC system and benchmark configurations are implemented using the RTI-MPC formulation described in Section~\ref{Sec: Controller_Implementation}.  
The same prediction models 
\eqref{Eq: discrete Elec_mdl}\text{--}\eqref{Eq: discrete therm mdl} and \eqref{Eq: RTI_linearised_coolant}, prediction horizon, objective function, constraints, and observer are employed across all configurations; they differ only in the admissible set of control inputs. For each benchmark, only the cooling channels corresponding to a given configuration are assumed physically available, while the total coolant flow rate remains identical across all configurations. Accordingly, for SC and bTC, only the lateral surface duty-cycle input $u_{\text{mod},s}$, and the bottom tab input $u_{\text{mod},b}$ are available, respectively. For bTSC, the admissible inputs are  $u_{\text{mod},b}$ and $u_{\text{mod},s}$, while for btTC, $u_{\text{mod},t}$ and $u_{\text{mod},b}$ are available. For both the proposed ITSC strategy and ES benchmark, all three inputs $u_{\text{mod},s},u_{\text{mod},t}$ and $u_{\text{mod},b}$ are available. In the ES configuration, the inputs are fixed according to $u_{\text{mod},s} = u_{\text{mod},t} = u_{\text{mod},b} = \frac{1}{3}$, resulting in an equal allocation of the total volumetric coolant flow rate among the three channels.

\begin{figure*}[tbh]
\centering
\includegraphics[width=0.9\linewidth]{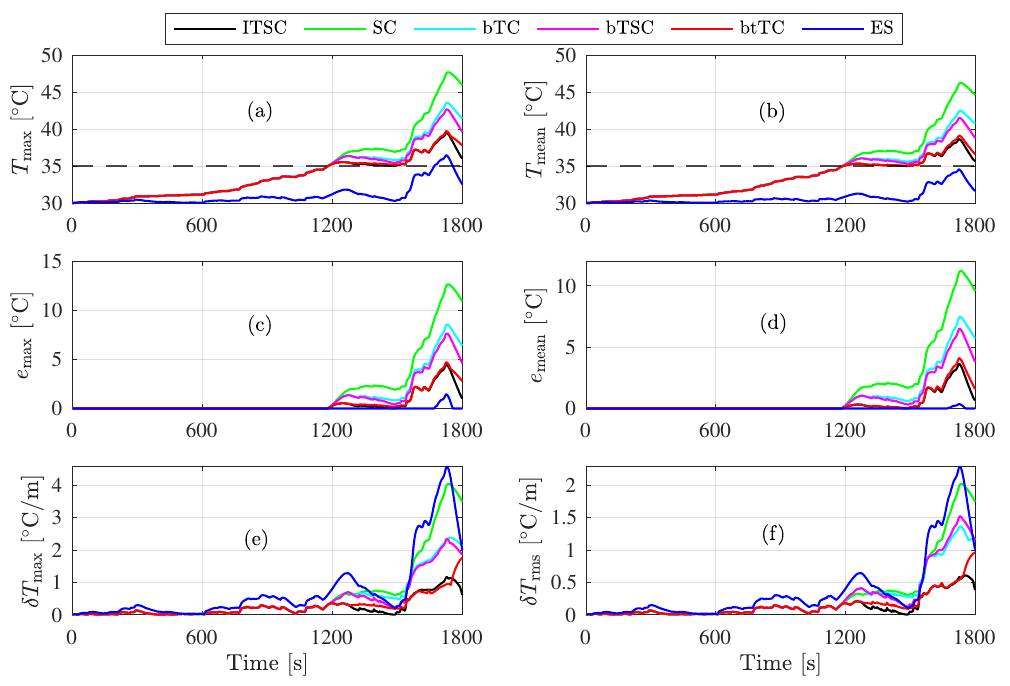}
\caption{Comparison of the proposed ITSC system and benchmark cooling configurations under the mWLTP. (a) and (b) present the maximum and mean temperature responses; (c) and (d), the maximum and mean tracking errors; and (e) and (f), the maximum and  RMS thermal gradients.}
\label{Fig: PerfMetrics}
\end{figure*}

\begin{figure}[tbh]
\centering
\includegraphics[width=1.0\linewidth]{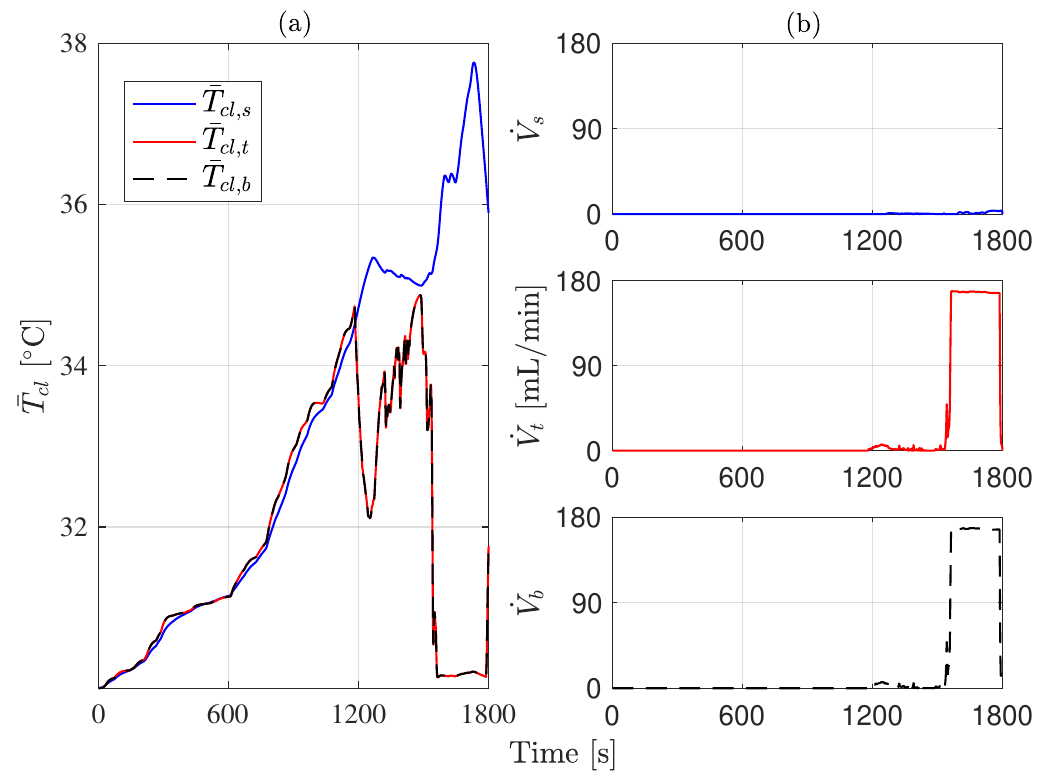}
\caption{ITSC system channel responses. (a) Mean coolant temperature evolution in the surface, top, and bottom channels. (b) Volumetric flow rate allocation.}
\label{Fig: Tcl_dotV}
\end{figure}

\subsection{ITSC against conventional cooling configurations} \label{}
The thermal performance of the proposed ITSC system relative to the benchmark cooling configurations are shown in Fig.~\ref{Fig: PerfMetrics}. 
From Figs.~\ref{Fig: PerfMetrics}(a)--(b), the ES configuration yields the lowest $T_{\max}$ and $T_{\mathrm{mean}}$ values. This behaviour is expected because ES continuously distributes the full available coolant flow across all three channels, thereby maximising the effective heat transfer area and total heat removal. However, this does not necessarily imply superior feedback regulation. ES operates in open-loop and cannot track a desired temperature reference. The deviation errors in Figs.~\ref{Fig: PerfMetrics}(c)--(d) provide a more direct assessment of temperature regulation relative to $T^r$. Although the temperature trajectory under ES remains below the reference (zero error) for most of the cycle, it does not track $T^r$ in any meaningful sense. It operates without adaptation to the instantaneous thermal demand and continuously uses the full pump capacity. Therefore, its lower temperature rise and tracking error are achieved at the expense of unnecessary coolant usage under moderate thermal loading.

Among the feedback-controlled configurations, the proposed ITSC system achieves the smallest reference overshoot. This indicates that the ITSC controller is able to exploit all available cooling areas to keep the cell temperature closer to the desired operating range. 
In contrast, SC exhibits the largest deviation error, indicating that lateral surface cooling alone is insufficient to regulate 
the thermally dominant internal region of the cell. 
Although bTC uses a smaller heat-transfer area than SC, 
it interestingly achieves lower temperature rise and smaller deviation above the reference. This is attributed to the homogenising effect of tab cooling: by reducing internal temperature gradients, the optimiser drives spatial temperatures closer together, which
indirectly constrains the peak temperature. The addition of both tab and surface cooling in bTSC leads to a further temperature reduction, while btTC improves the response further still. These trends highlight the importance of tab cooling for extracting heat from the cell more effectively than surface cooling alone.

Figs.~\ref{Fig: PerfMetrics}(e)--(f) show that the proposed ITSC system yields the lowest thermal gradients among all configurations. The improvement over SC, bTC, and bTSC is substantial, while btTC is the closest conventional competitor. The flow allocation in Fig.~\ref{Fig: Tcl_dotV}(b) provides useful insight into this behaviour: the ITSC controller predominantly routes coolant through the top and bottom channels during periods of high thermal loading. 
Under the chosen cost weights ($\lambda_1 = \lambda_2$), the optimiser assigns comparable importance to temperature tracking and gradient minimisation. Since tab cooling is intrinsically more effective at homogenising the temperature field, the controller naturally favours the tab channels under this tuning. 
The btTC configuration exclusively utilises the top and bottom axial paths, thus similar gradient response arises. However, ITSC retains the surface channel as an additional controllable degree of freedom, allowing the optimiser to balance gradient suppression and temperature tracking more flexibly.

Across all feedback-controlled configurations, substantial coolant-flow actuation commences around $t=1500$~s, coinciding with the onset of sustained high heat generation in the latter phase of the mWLTP. Prior to this period, the optimiser assigns little or no coolant flow because the cell temperature remains close to or below the reference. This is consistent with the one-sided deviation metric and the cooling-only nature of the actuation. Coolant flow is primarily used to suppress overshoot above $T^r$, rather than to force unnecessary cooling when the cell is already within the desired operating range. 

Taken together, these results show that the proposed ITSC system provides the best overall 
trade-off among the considered configurations. While ES attains the lowest temperature rise through continuous full-flow operation, it lacks feedback adaptation and does not optimise coolant usage relative to the reference. ITSC, by contrast, actively balances temperature regulation 
and thermal gradient minimisation under the same flow budget. A quantitative summary of the performance metrics is reported in Table~I.

\begin{table}[tbh]
  \caption{Maximum values of thermal performance metrics under the mWLTP for the proposed ITSC system and benchmarks.}
 \vspace{-0.5cm}
 \begin{center}
\begin{tabular}{c|c|c|c|c|c|c}
\hline
   Configurations 
   & $T_{\max}$ 
   & $T_{\text{mean}}$ 
   & $e_{\max}$ 
   & $e_{\text{mean}}$ 
   & $\delta T_{\max}$ 
   & $\delta T_{\text{mean}}$ 
  \\
 \hline
    ITSC    &   39.47    &   38.65    &   4.47    &   3.65    &   \bf{1.17}    & \bf{0.61} \\
    SC      &   47.66    &   46.22    &   12.66   &   11.22   &   4.05         &  2.02     \\
    bTC     &   43.55    &   42.47    &   8.55    &   7.47    &   2.38         &  1.36     \\
    bTSC    &   42.65    &   41.50    &   7.65    &   6.50    &   2.34         &  1.52     \\
    btTC    &   39.71    &   39.10    &   4.71    &   4.10    &   1.76         &  0.96     \\
    ES      & \bf{36.45} & \bf{34.54} & \bf{1.45} & \bf{0.36} &   4.57         &  2.28     \\
\hline
    \end{tabular}
     \label{Tab: Summary}
  \end{center}
\end{table}

The mean coolant temperature evolution in the three ITSC channels is shown in Fig.~\ref{Fig: Tcl_dotV}(a). The surface channel exhibits the highest coolant temperature, compared to the top and bottom channels. This follows directly from the flow allocation in Fig.~\ref{Fig: Tcl_dotV}(b), where the controller assigns only a small fraction of the total coolant flow to the surface channel and substantially larger flow rates to the top and bottom channels. This trend agrees with the coolant energy balance in \eqref{Eq: coolant_dynamics}. Defining the characteristic coolant residence time as $\tau_{\mathrm{res}} = V_c/\dot{\mathcal{V}}$, the advective exchange with the inlet scales as $\dot{\mathcal{V}}/V_c = 1/\tau_{\mathrm{res}}$. At lower flow rates, the residence time increases and the advective replacement of warmed coolant by colder inlet coolant weakens. As a result, the coolant temperature rises toward the local battery temperature $T_{bc}$ through the convective heating term. 
Conversely, at higher flow rates, advective transport dominates and constrains the coolant temperature more closely to $T_{cl,\mathrm{in}}$. This increases the battery-to-coolant temperature difference and therefore enhances heat removal.

\section{Conclusion} \label{Sec: Conclusions}
This article has proposed an integrated tab-surface cooling (ITSC) system for battery thermal management, formulated as a receding-horizon optimal control problem in which coolant is dynamically allocated among the lateral surface and tab channels.
To enable this formulation, a first-principles coolant model with volumetric flow rate as the manipulated input was derived and coupled with battery thermal and electrical models. A valve-actuation model was also incorporated to capture actuator-level coolant routing. The resulting battery--coolant model served as the predictive model within an MPC framework targeting temperature regulation and thermal gradient minimisation. 

To reduce the prohibitive computational expense, a real-time iteration (RTI)-MPC was developed for the proposed coolant-allocation problem. 
By locally linearising the coolant dynamics and solving a single quadratic programme per sampling instant, the RTI-MPC closely reproduced the NMPC thermal response while reducing the online computation time well below the sampling interval. This demonstrates that the proposed RTI-MPC can retain near-NMPC thermal performance at a
computational cost compatible with real-time implementation. 

Evaluation results under a real-world driving profile showed that the ITSC system provides the
lowest thermal gradients among all considered conventional cooling configurations. Also, it achieved the lowest temperature tracking error among the feedback-controlled cooling configurations. The rule-based equal-split configuration attained the lowest tracking errors due to the continuous use of the full coolant flow capacity across all channels. This was, however, achieved without feedback adaptation and at the expense of larger thermal gradients. The ITSC system, by contrast, actively allocated coolant according to the instantaneous thermal demand, balancing temperature tracking and gradient minimisation under the same total flow budget.

Future work will address joint pump-flow and coolant-allocation control, where the total coolant flow rate is treated as an additional decision variable.
 Further extensions will incorporate explicit degradation- and energy-aware cost formulations to quantify long-term benefits, and will apply the proposed framework to module- and pack-level assemblies where inter-cell thermal interactions and coolant-network constraints become significant.

\appendices
\section{Equivalent circuit model matrices} \label{App: ECM}
This appendix provides the explicit expressions of the system matrices associated with ECM \eqref{Eq: ECM}. 

\begin{subequations}
\label{Eq: ECM_Matrices}
\begin{align}
\mathcal{A}_e &= \begin{bmatrix}
0 & 0 & \cdots & 0 \\
0 & \frac{-1}{R_1 C_1} & & 0 \\
\vdots & & \ddots & \vdots \\
0 & 0 & \cdots & \frac{-1}{R_p C_p}
\end{bmatrix}, \:
\mathcal{B}_e = \begin{bmatrix}
\frac{-1}{3600C_{nom}} \\ \frac{1}{C_1} \\ \vdots \\ \frac{1}{C_p}
\end{bmatrix}, \\
\mathcal{C}_e &= \begin{bmatrix}
0 & -1 & \cdots & -1
\end{bmatrix}, \quad \quad\:\:
D_e = -R_0,
\end{align}
\end{subequations}
where $C_{nom}$ is the nominal charge capacity of the cell. 

\section{Derivation of the mean coolant temperature dynamics}
\label{App: coolant}

This appendix provides the detailed derivation of the coolant temperature dynamics presented in \eqref{Eq: coolant_dynamics}.
We apply the first law of thermodynamics to a control volume consisting of the coolant in thermal contact with the battery surface. The overall energy balance is given by
\begin{equation}
\label{Eq: coolant_energy_balance_app}
\frac{dE}{dt} = Q_{\text{conv}} + \dot{E}_g + \dot{E}_{\text{in}} - \dot{E}_{\text{out}} - \dot{W}_f,
\end{equation}
where $E$ is the total energy within the control volume, $Q_{\text{conv}}$ is the convective heat transfer from the battery surface to the coolant, $\dot{E}_g$ is the heat generated within the coolant, $\dot{E}_{\text{in}}$ and $\dot{E}_{\text{out}}$ are the energy fluxes at the inlet and outlet, respectively, and $\dot{W}_f$ is the flow work rate.

Under the stated simplifications, the stored energy reduces to sensible thermal energy, yielding
\begin{equation}
\label{Eq: sensible_energy}
\frac{dE}{dt} = m_c c_{p,c} \frac{d \bar{T}_{cl}}{dt},
\end{equation}
where $m_c$ is the coolant mass in the control volume.

The convective heat transfer is given by Newton's law of cooling:
\begin{equation}
\label{Eq: q_conv_app}
Q_{\text{conv}} = h A_{ht}(T_s - \bar{T}_{cl}).
\end{equation}
Since the coolant is passive, no internal heat generation occurs, i.e., $\dot{E}_g = 0$.

The energy transport due to fluid motion is expressed using the specific enthalpy formulation. For a mass flow rate $\dot{m}_c$, the net energy flow becomes
\begin{equation}
\label{Eq: enthalpy_balance}
\dot{E}_{\text{in}} - \dot{E}_{\text{out}} = \dot{m}_c c_{p,c} (T_{cl,\text{in}} - T_{cl,\text{out}}).
\end{equation}
Assuming a lumped control volume, the outlet temperature is approximated by the mean coolant temperature, i.e., $T_{cl,\text{out}} \approx \bar{T}_{cl}$. Neglecting mechanical energy contributions and pressure work, substituting \eqref{Eq: sensible_energy}--\eqref{Eq: enthalpy_balance} into \eqref{Eq: coolant_energy_balance_app} yields
\begin{equation}
\label{Eq: intermediate_balance}
m_c c_{p,c} \frac{d\bar{T}_{cl}}{dt} = h A_{ht}(T_s - \bar{T}_{cl}) + \dot{m}_c c_{p,c}(T_{cl,\text{in}} - \bar{T}_{cl}).
\end{equation}
Finally, expressing the coolant mass and mass flow rate in terms of density and volumetric flow rate, $m_c = \rho_c V_c$ and $\dot{m}_c = \rho_c \dot{\mathcal{V}}$, yields
\begin{equation}
\label{Eq: final_coolant_app}
\frac{d\bar{T}_{cl}}{dt} = \frac{h A_{ht}(T_s - \bar{T}_{cl})}{\rho_c V_c c_{p,c}} + \frac{\dot{\mathcal{V}}(T_{cl,\text{in}} - \bar{T}_{cl})}{V_c},
\end{equation}
which is identical to \eqref{Eq: coolant_dynamics}.

\section{Determining convection coefficients} \label{App: h_coeff}
To determine the appropriate convection coefficients associated with each cooling channel, we utilise the general relation in \eqref{Eq: h_general}. The hydraulic diameter $D_h$ depends on the specific geometry of each channel.

For the lateral surface, the coolant flows through a narrow annular channel of height $L$ and thickness $\epsilon_s$. Approximating the local geometry as a rectangular duct, the hydraulic diameter is given by
\begin{equation}
\label{Eq: Dh_s}
D_{h,s} = \frac{4A_{cf,s}}{P_s},
\end{equation}
where $A_{cf,s} = L\epsilon_s$ is the flow cross-sectional area and $P_s = 2(L + \epsilon_s)$ is the wetted perimeter.

For the top and bottom channels, the coolant flows between parallel plates of thickness $\epsilon_t$ and $\epsilon_b$, respectively. In this case, the hydraulic diameters are given by \cite{incropera1985fundamentals}
\begin{equation}
\label{Eq: Dh_t_b}
D_{h,t} = 2\epsilon_t, \quad D_{h,b} = 2\epsilon_b.
\end{equation}

Substituting \eqref{Eq: Dh_s} and \eqref{Eq: Dh_t_b} into \eqref{Eq: h_general}, the convection coefficients become
\begin{equation}
\label{Eq: hs_t_b}
h_s = \frac{\text{Nu} \cdot k_c (L + \epsilon_s)}{2L\epsilon_s}, \quad
h_t = \frac{\text{Nu} \cdot k_c}{2\epsilon_t}, \quad
h_b = \frac{\text{Nu} \cdot k_c}{2\epsilon_b}.
\end{equation}

While $\epsilon_t$ and $\epsilon_b$ are often equal, the known higher heat generation from the positive tab areas compared to the negative \cite{an2018modeling,lyu2023experimental} may necessitate designing $\epsilon_t \neq \epsilon_b$ to accommodate the resulting thermal asymmetry. $D_h$ can also serve as a design parameter 
to guide the optimal sizing of the cooling channels.

\bibliographystyle{IEEEtran}%
\bibliography{references}%

\end{document}